# Sparse movement data can reveal social influences on individual travel decisions

Tyler R. Bonnell[1], S. Peter Henzi[1,2], and Louise Barrett[1,2]

[1]Department of Psychology, University of Lethbridge, Alberta, Canada

[2]Applied Behavioural Ecology and Ecosystems Research Unit, University of South Africa

**Abstract:**

The monitoring of animal movement patterns provides insights into animals' decision-making behaviour. It is generally assumed that high-resolution data are needed to extract meaningful behavioural patterns, which potentially limits the application of this approach. Obtaining high-resolution movement data continues to be an economic and technical challenge, particularly for animals that live in social groups. Here, we test whether accurate movement behaviour can be extracted from data that possesses increasingly lower temporal resolution. To do so, we use a modified version of force matching, in which simulated forces acting on a focal animal are compared to observed movement data. We show that useful information can be extracted from sparse data (i.e., collected over minutes instead of seconds). We apply this approach to a sparse movement dataset (average revisit time of 9min) collected on the adult members of a troop of baboons in the DeHoop Nature Reserve, South Africa. We use these data to test the hypothesis that individuals are sensitive to isolation from the group as a whole or, alternatively, whether they are sensitive to the location of specific individuals within the group. Using data from a focal animal, our data provide support for both hypothesis, with stronger support for the latter. Our focal animal showed consistent patterns of movement toward particular individuals when distance increases beyond 5.6m. Although the focal animal was also sensitive to the group, this occurred only on a small number of occasions (4.2% of the data) when the group as a whole was highly clustered as a single entity away from the focal animal. We suggest that specific social interactions may thus drive overall group cohesion. Given that sparse movement data is informative about individual movement behaviour, we suggest that both high (~seconds) and relatively low resolution (~minutes) datasets are valuable for the study of how individuals react to and manipulate their local social and ecological environments.

1. **Introduction:**

Fine-grained analysis of the movement strategies of mobile animals have benefited enormously from technical advances in global positioning systems (GPS). Increased miniaturization, higher frequency captures rates, and longevity of battery life all permit the collection of rich datasets from animal-mounted GPS units (Nathan *et al.* 2008; Cagnacci *et al.* 2010). In association with the development of new analytical techniques (Laube, Imfeld & Weibel 2005; Gurarie, Andrews & Laidre 2009; Dalziel *et al.* 2015; De Groeve *et al.* 2015), such data are beginning to provide answers to long-standing questions in movement ecology, and are driving new research programs, particularly with respect to social animals (Hebblewhite & Haydon 2010; Lukeman, Li & Edelstein-Keshet 2010; Krause *et al.* 2013; Kays *et al.* 2015; Strandburg-Peshkin *et al.* 2015).

There are, however, some limitations associated with the use of animal-attached GPS devices, not all of which can be addressed through technical developments. While battery life may improve, allowing more than short high-resolution snapshots of activity (Strandburg-Peshkin *et al.* 2015), equipment failure will persist. Moreover, the need to attach devices to the animals can raise ethical, logistical and technical issues (Handcock *et al.* 2009). There are also economic costs that may limit access to some researchers, or require sampling of only a very few individuals. Even in cases where one can capture all members of a social group, repeated immobilization to affix and retrieve collars and data may limit researchers' ability to rely on these techniques alone.

Many social animals, most notably primates, can be habituated and followed sufficiently closely to allow human observers to mimic automatic spatial data collection with hand-held GPS dataloggers (Sugiura, Shimooka & Tsuji 2011; Aureli *et al.* 2012; Heesen *et al.* 2015). This method reduces costs and may lift other logistical constraints, but the resolution of the data (i.e., number of records per unit time) is far lower than that deliverable by fully automated, animal-mounted techniques. Nevertheless, if such data can be shown to produce reliable spatio-temporal movement patterns, then low-resolution approaches may offer a sustainable, flexible and reliable alternative approach to automated methods.

Here, we address the extent to which the individual actions that underpin observed collective motion can be detected as the temporal resolution of data is lowered. We implement a modified version of Eriksson et al.'s (2010) force matching method. This sets behavioural rules describing motion, and then fits parameters to these rules that minimize the deviation between observed and predicted movements. These rules are used both to describe those elements of the environment to which an individual is sensitive (i.e., will respond to by moving) and the kinds of behavioural response elicited.

To test how the temporal resolution of data affects the reliability of the movement trajectories, we use agent-based modeling to specify the rules governing individual movement in a simulated group of agents. We specify these rules based on the empirical findings described by Strandburg-Peshkin (2015), obtained using high-resolution data from animal-attached GPS units. We record the resultant trajectories of our simulated agents based on similar high-frequency sampling, and then subsample these trajectories to generate datasets of differing temporal resolution. We then apply force matching to each of these, and assess the extent to which the results obtained are able to identify the rule we had built into the agents. In this way, we can assess the extent to which our modified force matching approach is able to: 1) identify the true conditions under which a given behaviour occurs, and 2) accurately identify the correct behaviour as temporal resolution declines. As a second step, we apply this approach to real-world movement data collected from a wild troop of baboons (*Papio hamadryas ursinus*) to test whether we are similarly able to detect group-following in a low-resolution dataset, and whether we detect the same patterns as Strandburg-Peshkin (2015).

Finally, we offer a "proof of concept" illustration of the way in which our method enables researchers to test between different hypotheses of group movement. Current theories of group movement are based on both leadership and consensus-formation within the group (Conradt &

Roper 2005; King *et al.* 2008). In baboons, for example, it has been shown that individuals are sensitive to the number of initiators and their agreement (Strandburg-Peshkin *et al.* 2015). There is also evidence to suggest that individuals are more likely to follow those with whom they share close social affiliations (King *et al.* 2011). These potential rules of thumb, e.g., follow close social affiliates or follow the majority of the group, are not necessarily mutually exclusive. This presents a real challenge when attempting to identify or differentiate between the possible mechanisms behind observed motions. For example, in the case of two clusters of initiators within a group, a larger group of initiators may be more likely to contain close social affiliates than a smaller group of initiators. This might lead to the observation that, on average, individuals move more toward the majority when, in fact, patterns of affiliation between individuals might be responsible. Using data from a focal individual, we therefore test the extent to which the movement of the focal animal is 1) sensitive to the spatial position of the group as a whole, and 2) sensitive to the spatial positions of particular individuals within the group, and whether it is possible to assess which is the most influential.

2. **Materials and Methods:**

*2.1. Modified force matching for sparse datasets*

The force matching method proposed by Eriksson et al. (2010) identifies optimal models of interactions in animal groups by adapting the method used to describe interactions among particles (Ercolessi & Adams 1994). In essence, it develops models that describe how other group members (hereafter 'associates') influence a focal animal, and then compares how well these models match the observed motion of the focal animal. As the method deals with measures of force (force = mass * acceleration), it relies on knowing the acceleration of each individual at each time-point in order to calculate the forces exerted on any one animal by its associates.

In sparsely measured movement data, acceleration cannot be estimated accurately, which presents a problem in applying the force matching method to coarse-grained datasets (Eriksson et al. 2010). If we modify our predictions, however, from those dependent on acceleration to those based solely on the direction of travel - a measure easily estimated for sparse movement data – it is possible to apply a version of force matching. This allows us to quantify the influence that others have on the direction of travel of a focal individual (Fig.1). It is then possible to test models that describe how the specific locations or movements of others influence the movement of a focal animal. This allows us to address questions, for example, about whether individuals show signs of moving to the centre of the group or whether they avoid higher-ranking individuals.

Specifically, our proposed procedure follows three steps: 1. identify a focal individual and, at each observed point $(x_t, y_t)$, measure its observed direction of travel $((x_{t+1}, y_{t+1})-(x_t, y_t))$ and the direction of travel from the preceding observation $((x_t, y_t)-(x_{t-1}, y_{t-1}))$; 2. use linear interpolation between sequential observations of all associates to estimate their position and direction of travel at the point of observation of the focal animal (Fig. 1). By using this approach we generate a dataset containing, the observed direction of travel associated with a given spatial structure of associates. It is then possible to 3. use this spatial structure to search for a model that minimizes the squared difference between the observed and predicted direction of travel: $\varepsilon = (\theta_{obs} - \theta_{pred})^2$.

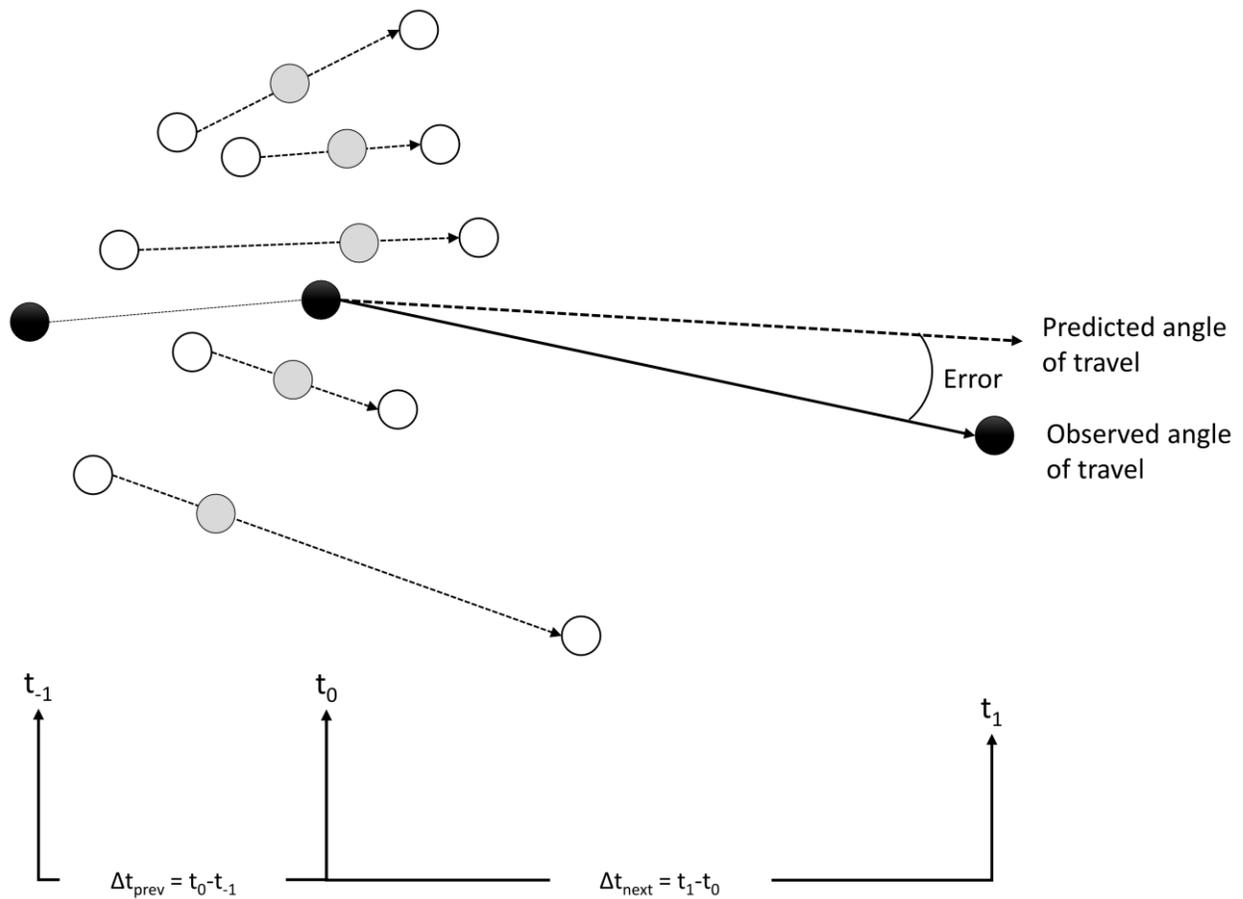

*Figure 1: Diagram of data extraction method for coarse data. The focal animal's location is indicated by the black circle, while grey circles identify the estimated position of other group members at the same observation time. The observed angle of travel ($\theta_{obs}$) between $t_0$ and $t_1$ is then contrasted with the modeled angle of travel ($\theta_{pred}$).*

There are three sources of error associated with this approach that increase as the data become more sparse. First, when the time elapsed between the current and succeeding observations increases ($\Delta t_{next}=t_0-t_1$), the estimated direction of travel will be influenced increasingly by other events that occur between the two time points, rather than only the group structure at time t. Second, as the time elapsed between preceding and current observation points of the focal individual increases ($\Delta t_{prev}=t_0-t_1$), the previous bearing will have an increasingly smaller influence on the subsequent one. The third source of error results from the interpolation of associates' positions and directions of travel, as the error of both estimates will increase as data become more sparse. Here, we make the assumption that these errors are unbiased (i.e., mean error =0), and suggest that, with sufficient data, consistent patterns can be extracted from the noise.

*2.2. Simulated dataset*

To test the feasibility of the modified force matching approach we used Repast Simphony (North *et al.* 2013) to simulate a group of agents (N=14) with fixed behavioural rules (i.e., known behaviour). The model was based on the findings of (Strandburg-Peshkin *et al.* 2015), who used a high-resolution dataset (every second), in which stationary baboons showed a sensitivity to following moving animals, "initiators", if there were many of them and they were in high directional agreement. The authors quantified directional agreement of the initiators as one minus the total circular variation of the angles from a focal individual to each initiator (min=0, max=1) (Fig. 2). To mimic these findings, we developed a simulation model of a group of agents with simple foraging behaviours. These agents actively search/move for food patches located on a 2D landscape, and maintain cohesion by moving towards the group once an isolation threshold has been reached (Fig. S1). This isolation threshold was based on the magnitude of the directional agreement (DA), and a sum of inter-individual distances (IID) to all associates. DA measures the relative clustering of associates in terms of direction from the focal animal, and the IID threshold captures the relative spread of the group from the focal animal (Fig. 2). Agents would thus surpass the isolation threshold when the group was spread out and their directional clustering was high. Specific values for defining this threshold were chosen to produce active movement in the simulated group (IID > $\alpha_{iid}$ = 350 m, DA > $\alpha_{da}$ = 0.8) (Fig. 2). This simulation generated movement data in which behaviour was nonlinear: foraging individuals move in a straight line or towards food patches when these are available, and only move based on the position of other group members when the isolation threshold is met. We applied this method to simulate a group of 14 individuals for a period of 48 h, recording every agent's position at one-second intervals.

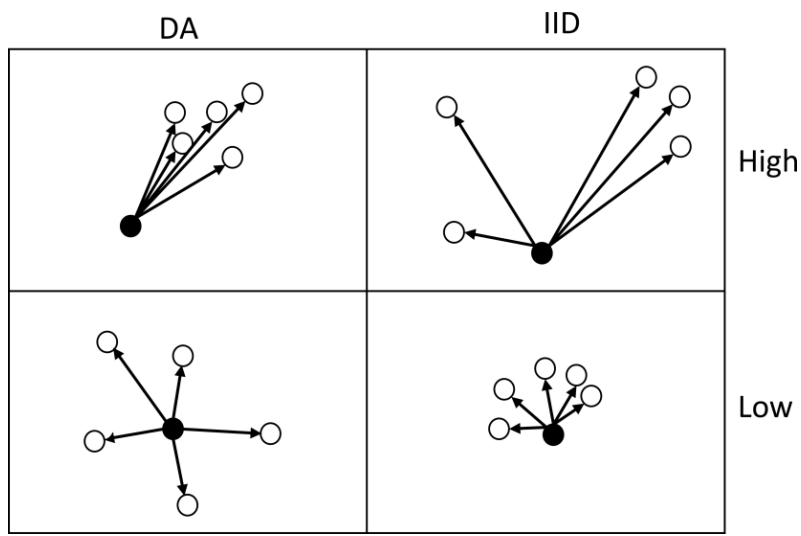

*Figure 2: Diagram of directional agreement (DA) and total inter-individual distance (IID). Both measures are egocentric, focusing on a focal animal.*

We generated additional datasets by progressively degrading the simulated data by sampling from a distribution of increasingly sparse revisit times. We used our own observed data on the movement patterns of baboons (see below) as a base estimate of the general structure of revisit times (i.e., the time elapsed between data points recorded for a given animal) captured in the field by a single observer, where the best fit distribution by maximum likelihood was lognormal with shape = 6.1 and scale = 0.6. We altered this distribution by shifting the shape parameter to produce a range of revisit time distributions, varying from more to less frequent revisits (i.e., mean revisit times of 1, 5, 10, 15, 20, 30, 40 min). Using these probability distributions, we resampled from the full dataset (1sec) to generate datasets of increasingly dispersed observations, mimicking variation in capture times (Fig. 3), containing respectively:

172775 (1sec), 3247 (1min), 610 (5min), 323 (10min), 212 (15min), 151 (20min) observations. Additionally, we quantified the effect of increasing the duration over which observations were recorded (referred to as the 'extent' of the data below) to assess whether increasing sample size in this way could compensate in any way for reduced data resolution. To do so, we simulated additional movement datasets to provide 48h, 96h, 144h, 192h of movement data with a temporal resolution of 10 min mean revisit times, containing respectively 323 (48h), 662 (96h), 967 (144h), 1310 (192h) observations.

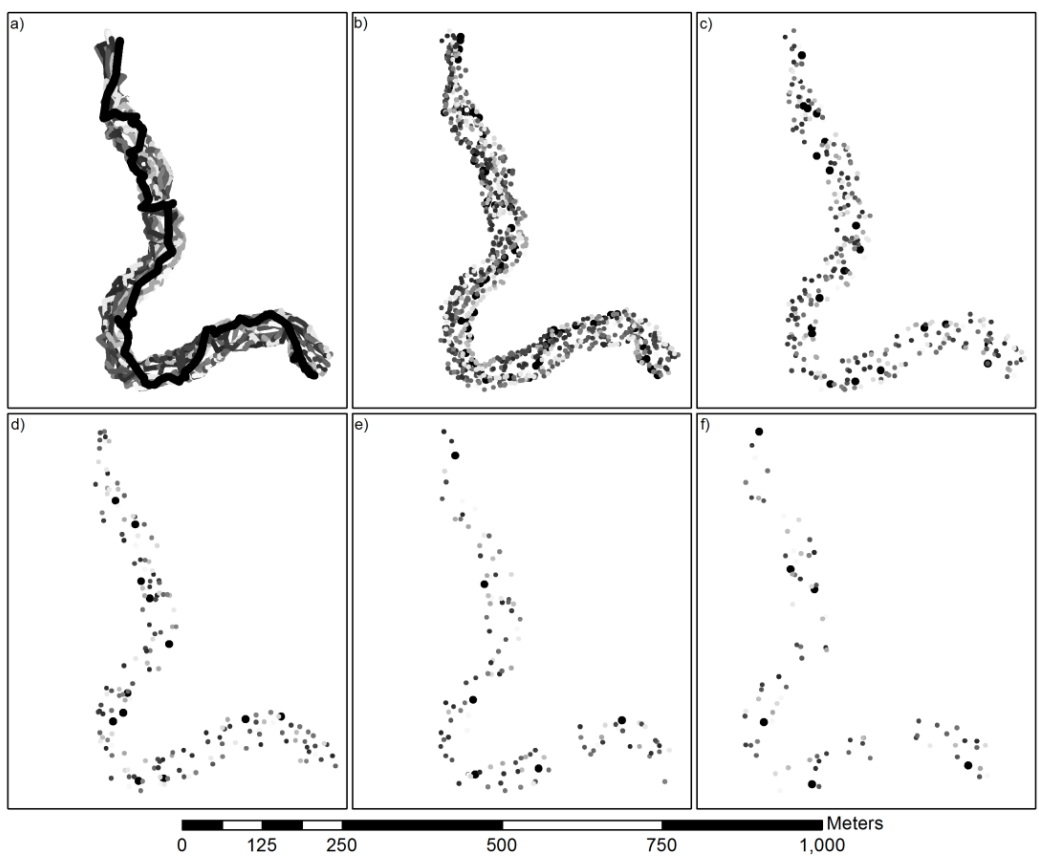

*Figure 3: a) positions of simulated agents over a 100 min period, with increasingly sparse observations: b) 1min, c) 5min, d) 10min, e) 15min, f) 20min mean revisit times. The thick black line/point represents the position of the focal individual on which the force matching was applied, associates are assigned smaller points with grey shading.*

2.3. Baboon dataset

Our field dataset consisted of 74 days of full-day follows of a baboon troop at the De Hoop Nature Reserve in South Africa (Barrett et al. 2004). Individual GPS points of all adult group members (N=14) were collected continuously throughout the day by an observer walking repeatedly from one end of the group to the other (see Andrienko et al. 2013). This generated 61842 points, with a mean revisit time for each individual of 9min. For our sample focal animal from this group - AL, the 2$^{nd}$ highest ranking female - this provided 4998 observations over 708h.

## 3. Analysis:

### 3.1. Force models tested

We propose two models based on an individual's response to group isolation and social isolation respectively (eq.1, 2). These allows us to test two hypotheses concerning the way that animals are thought to maintain cohesion within a group: one in which the animal is largely responding to the group as a set of homogenous individuals, the other where individuals respond to the group as a heterogeneous collection of individuals. Eq. 1 assumes the motion of the focal animal is solely influenced by the group, whereas eq. 2 includes the influence of both the group as a whole and individuals within the group. By incorporating both hypotheses within eq. 2 we can compare between these two hypotheses.

$$\hat{v}_t = \beta_1 \hat{v}_{t-1} + (\beta_2 \hat{v}_{CM} | \text{IID}_t > \alpha_{iid}, \text{DA}_t > \alpha_{da}) \qquad \text{eq.1}$$

$$\hat{v}_t = \beta_1 \hat{v}_{t-1} + (\beta_2 \hat{v}_{CM} | \text{IID}_t > \alpha_{iid}, \text{DA}_t > \alpha_{da}) + \sum_{j=1}^{n} (\beta_{i,j} \hat{v}_{i,j} | D_{i,j,t} > \alpha_{sd}) \qquad \text{eq.2,}$$

where $\hat{v}_i$ is the observed travel direction of the focal animal *i*, and is influenced by external factors: $\hat{v}_{t-1}$ is the previous direction of travel (i.e. previous bearing), $\hat{v}_{CM}$ is the circular mean of the directions to all associates from the focal animal i, and $\hat{v}_{i,j}$ is the direction to individual j from focal animal *i* (Fig. 4). The parameters $\beta_1$, $\beta_2$ and $\beta_{i,j}$ represent the weights of each influencing factor, and each provides the strength of its predictive influence on the focal individual's direction of travel. Parameters $\alpha_{iid}$, $\alpha_{da}$, and $\alpha_{sd}$ are estimates of the conditions in which a focal individual is considered isolated, e.g., a highly spread group ($\alpha_{iid}$) in which directional agreement is high ($\alpha_{da}$), and/or the distance from preferred group associates is high $\alpha_{sd}$. We measured group isolation of the focal animal *i* in terms of directional agreement in associates (DA), and the spread of associates (IID) relative to the focal individual. Social isolation of the focal individual *i* was measured as the straight line distance between two individuals ($D_{i,j}$).

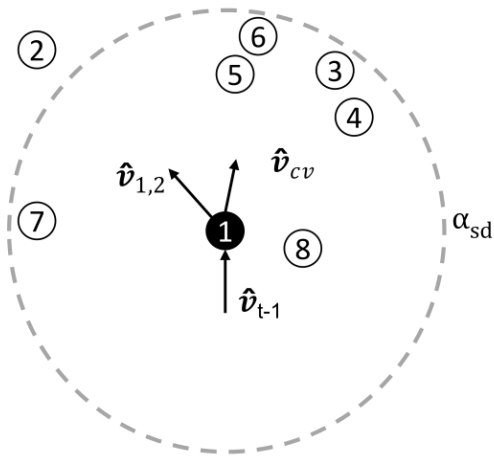

*Figure 4: Visual diagram of equation 2 applied to one focal baboon (id 1).*

## 3.2. Fitting force models

We used GeoTools in java ([www.geotools.org](www.geotools.org)) to develop a tool that extracts the direction of travel of a focal individual, the position and direction of travel of associates, the mean circular angle, and inter-individual distances at each observation time. The output from this tool can then be treated as an overdetermined system of equations, describing changes in x- and y-positions. In the case of linear equations, it is then possible to fit parameters by a least squares method ($X = (A^t A)^{-1} A^t Y$). Alternatively, with non-linear equations, such as eq.1 and 2, optimization methods can be used. We used the DEoptim algorithm (Mullen *et al.* 2011) in the R environment (R Core Team 2015) to minimize the sum of squares difference between observed and predicted direction of motion. As our equation is linear with conditional statements, we used the DEoptim algorithm to select parameters for the condition parameters ($\alpha_{iid}$, $\alpha_{da}$, and $\alpha_{sd}$). We then used non-negative least squares (Soetaert, Van den Meersche & van Oevelen 2009) to solve the remaining linear equation and return the sum of squares difference between observed and predicted directions of travel to the DEoptim algorithm. Non-negative least squares were used as we are interested only in attractive forces in our hypothesis. This optimization method reduces the parameter space explored by DEoptim (i.e., parameter space of 3), and takes advantage of the linear aspects of equation 1 and 2. For the simulated datasets, the range explored by DEoptim of the three conditional parameters was set at 0-1000 m for $\alpha_{iid}$, 0-1 for $\alpha_{da}$, and 0-10 m for $\alpha_{sd}$. For the baboon dataset, we kept the range of conditional parameters the same as in the simulated case, but increased the range for social distance ($\alpha_{sd}$) to 0-1000m, as we are less certain about the potential range of this parameter.

4. **Results:**

*4.1. Simulation results*

When the frequency of the simulated data was at the maximum resolution possible (i.e., 1 sec observations), the modified force matching approach successfully identified the influence of the group as a whole on focal animal movement, as opposed to the influence of individual associates (Fig. 5a), with estimated force coefficients of zero for all individual associates on the focal animal (Fig. 5a). Furthermore, the method was able to identify the correct conditional parameters defining group isolation: $\alpha_{iid} = 350.12, CI = (349.83, 350.39); \alpha_{da} = 0.80, CI = (0.801, 0.800))$. As we resampled the data to lower temporal resolutions, we quickly lost accuracy in the estimates of $\alpha_{iid}$, and $\alpha_{da}$ (Fig. 6). More promisingly, however, the ability to distinguish between our two competing hypotheses was much less sensitive to reductions in data frequency (Fig. 5): specifically, the estimated force coefficients for associates, i.e., effect size, were negligible compared to that of the group as a whole (Fig. 5).

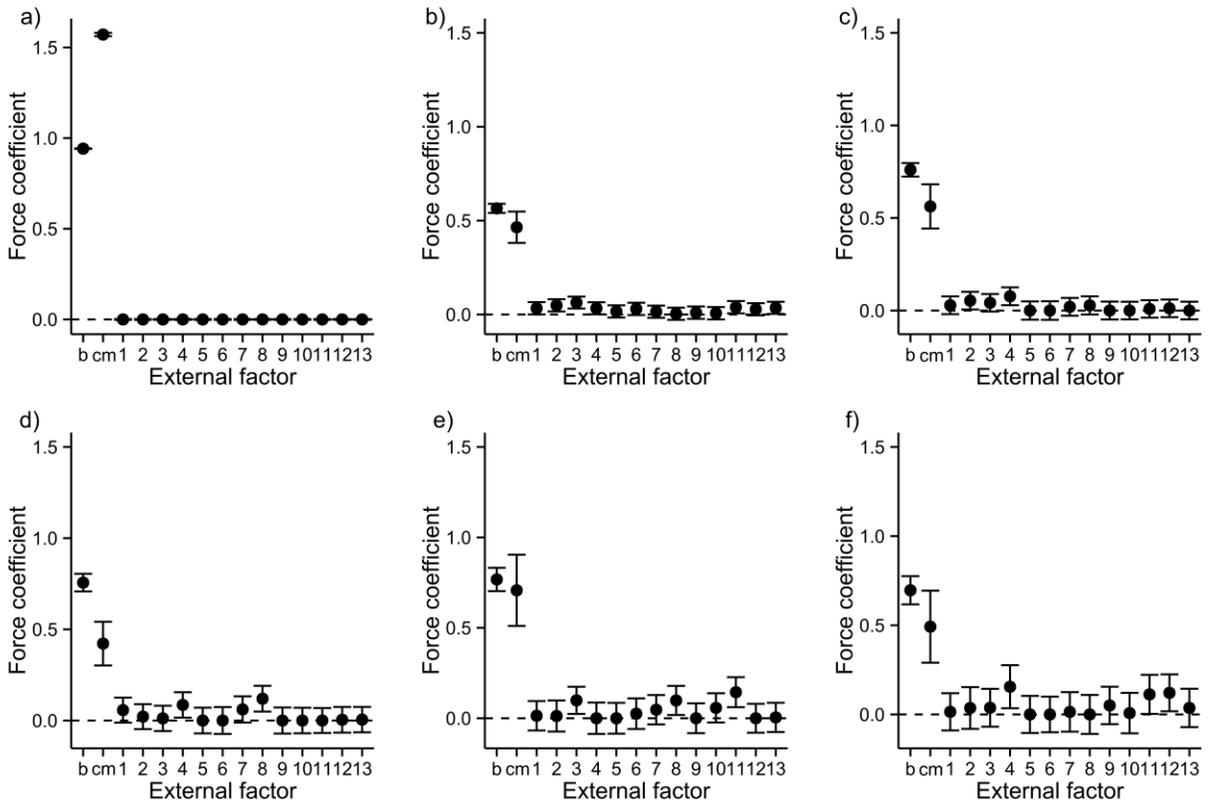

*Figure 5: Estimates of force coefficients for the a) non-reduced dataset, b) 1 min, c) 5 min, d) 10 min, e) 15 min, and f) 20 min mean revisit times. External factor "b" represents the force of the previous bearing, "cm" is the force towards the circular mean, and the numbered factors represent the pull to each group mate.*

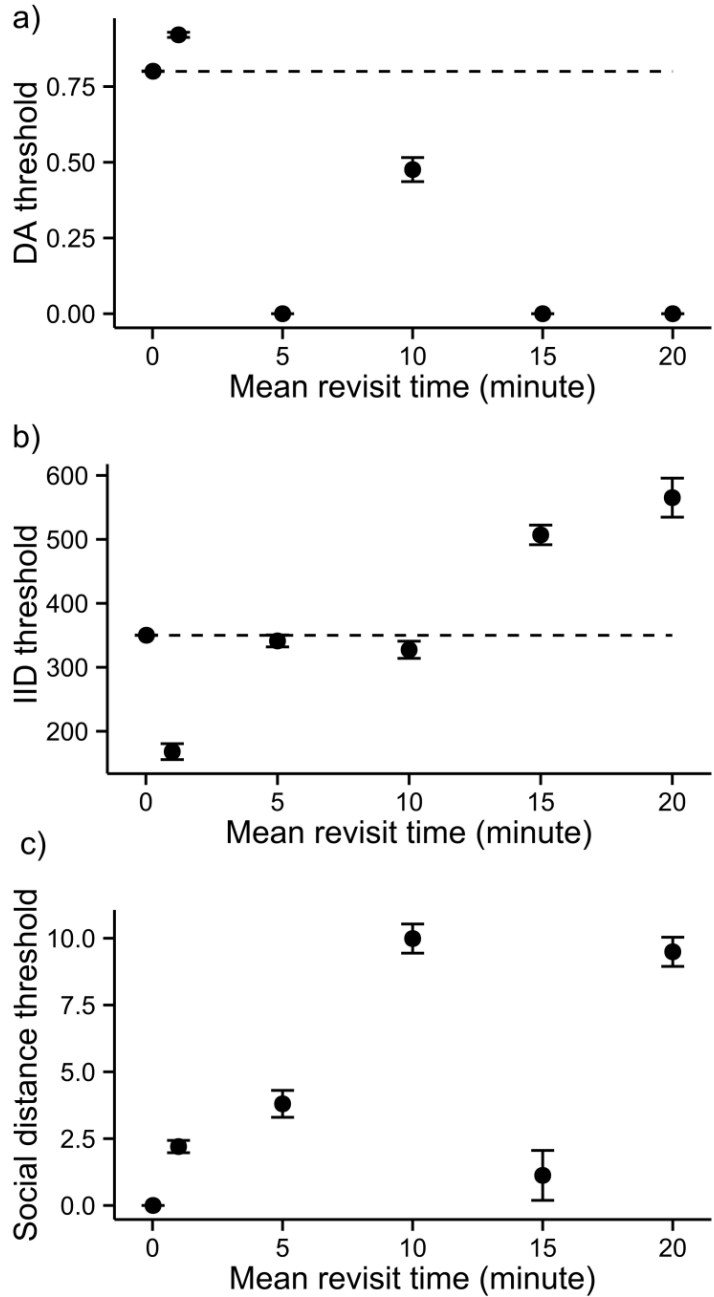

*Figure 6: Estimated conditional parameters as the temporal frequency of movement data is decreased: a) magnitude of directional agreement in associates, b) sum of the inter-individuals distances from the focal animal, c) minimum social distance, determining when a focal individual will follow. Error bars represent 95% confidence interval for each parameter estimate, dashed lines are the true values.*

When the time-frame over which data are analyzed is increased from 48h to 96h, 144h and 192h, we found that the estimates of force coefficients and conditional parameters became more accurate (Fig. 7). At 192h of observation, all associates force coefficients 95% confidence interval contains 0.0.

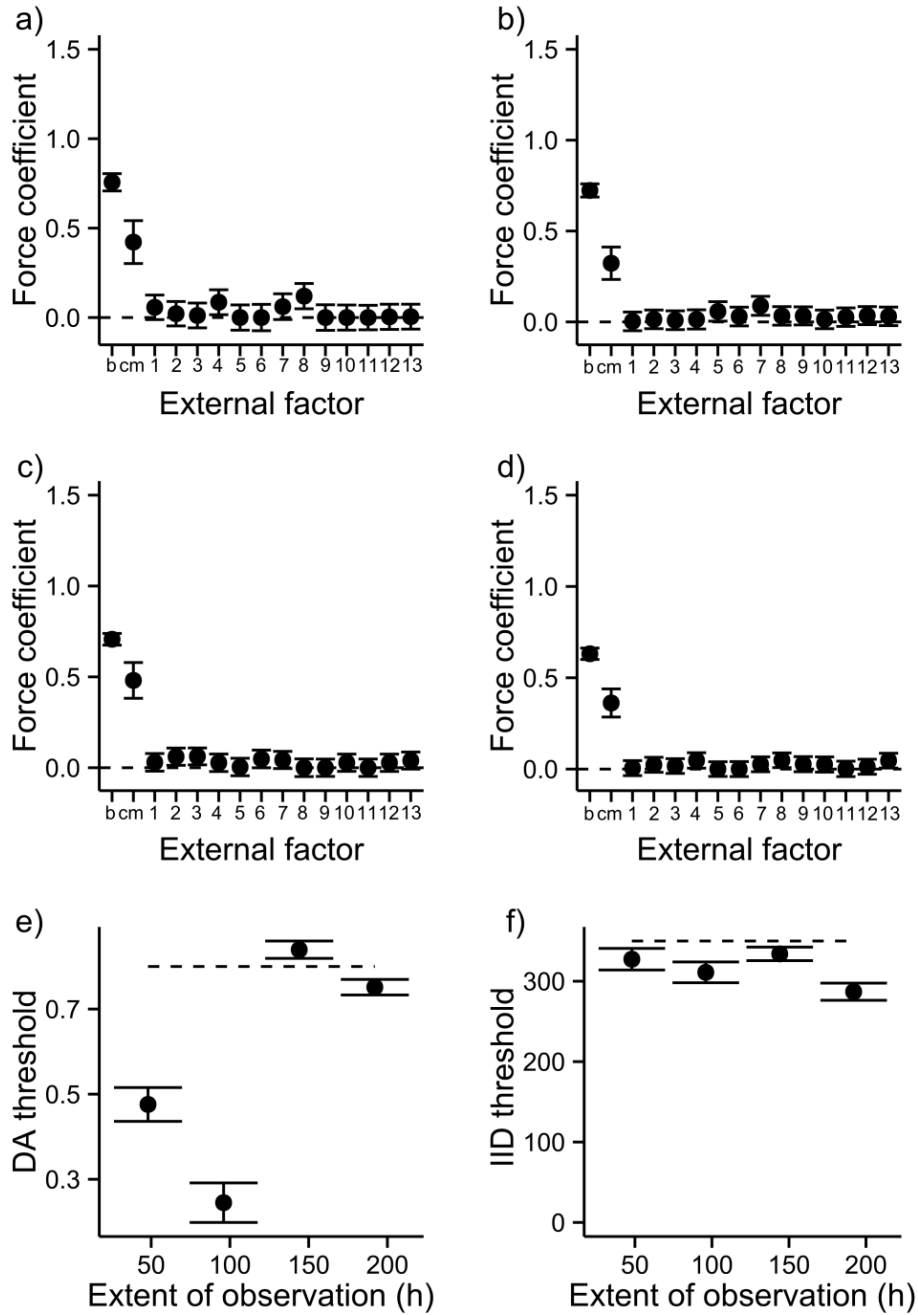

*Figure 7: Estimated force coefficients as the extent of data is increased from a) 48h, b) 96h, c) 144h, and d) 192h. Conditional parameter estimates are presented in e), f), and where the dashed horizontal lines represent the true value. Vertical bars are the 95% confidence intervals.*

*4.2. Baboon results*

Applying the modified force matching method to a sample focal animal in our real baboon data-set, using eq. 1, we found that AL was sensitive to the previous bearing ($\hat{v}_{t-1}$), and the position of the group as a whole ($\hat{v}_{cm}$). Our results are therefore comparable to those obtained with high-resolution data by (Strandburg-Peshkin *et al.* 2015), despite our lower temporal resolution. More specifically, the model suggested that, when directional agreement in the group was above 0.13 and the spread of the group was larger than 5.1m, the focal animal was influenced by mean group direction. These conditions apply in 96% of the observations, indicating a wide-spread influence of the group on travel direction.

When we applied eq. 2 to the baboon dataset in order to test between the two hypotheses of cohesion (sensitivity to the group, versus sensitivity to specific individuals), we found that AL's direction of travel showed signs of being sensitive to previous bearing ($\hat{v}_{t-1}$), the group as a whole ($\hat{v}_{cm}$), and specific individuals within the group ($\hat{v}_{i,j}$) (Fig. 8). The largest influence on direction of travel was that of the previous heading of the animal ($\hat{v}_{t-1}$). When we examine the estimated isolation conditions under which the $\hat{v}_{cm}$, and $\hat{v}_{i,j}$ influenced travel direction, we found that $\hat{v}_{cm}$ was estimated to become a factor only when the magnitude of the directional agreement (DA) was greater than 0.94, and that the measure of IID was not influential (95% confidence interval = 0, 15.6 m). The directional agreement condition (i.e., DA greater than 0.94) occurred only in 4.2% of the data, suggesting that moving toward the group as a whole was infrequent and occurred only when directional agreement was very high. In terms of moving towards particular animals in the group, AL was sensitive (in order of highest to lowest effect) to the highest ranking female (SA, id 9), the alpha male (SC, id 11), the third highest ranking female (VI, id 13), and a low ranking female (AC, id 1) (Fig. 9). She also showed signs of attraction toward a transient sub-adult male (KN, id 0). AL was estimated to be attracted to these specific individuals when their distance was over 5.6 m (95% confidence interval = 1, 10.2 m). All estimates of force coefficients to other animals in the group had 95% CIs that included 0.

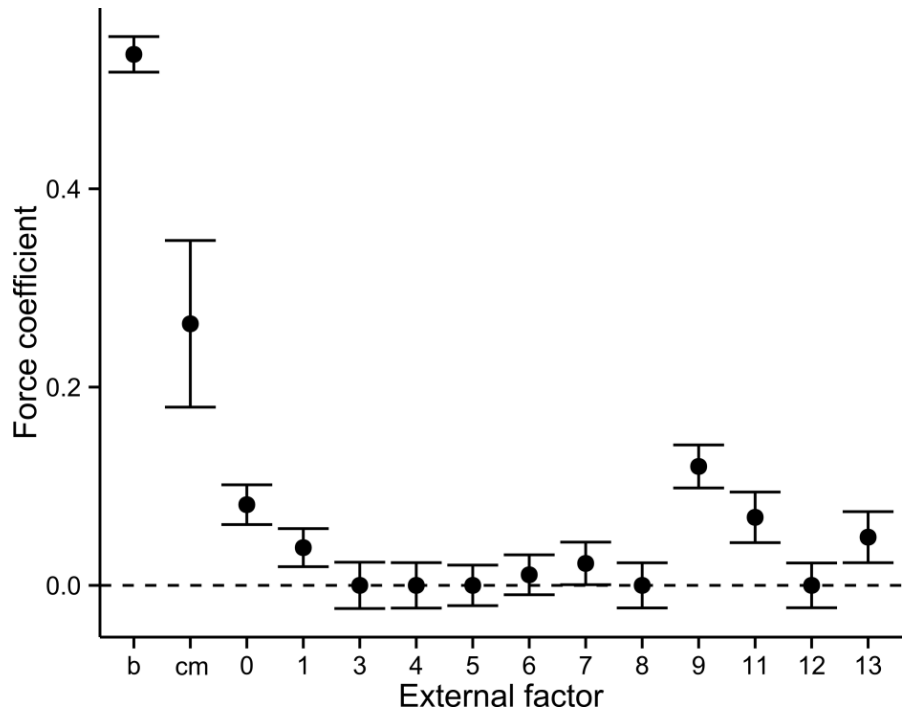

*Figure 8: Estimated force coefficients from one focal baboon in the observed GPS data. External factor "b" represents the previous bearing, "cm" represents the pull towards the mean circular direction of the group from the focal induvial, and the numbered factors represent the pull towards specific individuals in the group.*

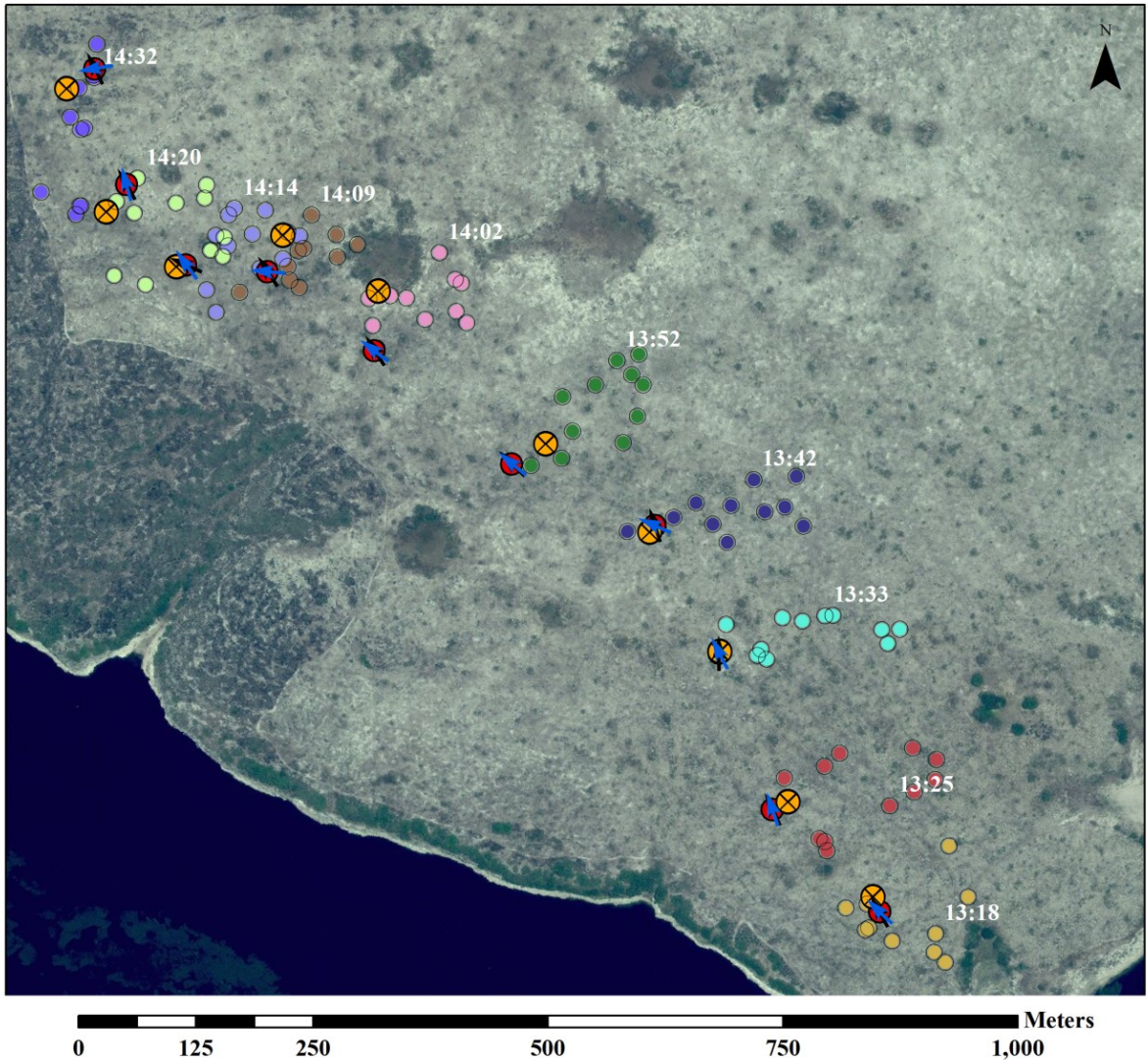

*Figure 9: Visual example of predicted vs. observed angle of travel. The focal animal is represented as a red circle, highest ranking female is represented as an orange circle with an "x", and the estimated position of all other group members are represented as smaller circles coloured based on the time of observation. The arrows superimposed on the focal animal, represent the observed direction of travel (blue) and the predicted direction of travel via eq.2 (black).*

Calculating the proportion of variation in direction of travel explained by each model, as one minus the residual sum of squares over the total sum of squares, we find that the r-squared of eq.1 was 0.30, whereas the r-squared of eq.2 was 0.32. The reduction of the influence of the group as a whole in eq.2, compared to eq.1, along with a similar $R^2$, suggests that the directional agreement of the group and the direction to specific individuals are, in fact, correlated. Indeed, the correlation between mean directional agreement of the group and the direction to the highest-ranking female (SA) relative to the focal animal was 0.59 (t=71.658, df=9610, p<0.001). Thus, socially influential individuals (e.g., those with many social partners) may be responsible for generating strong directional group movement. Our method does help to tease out these effects to some extent: AL was most influenced by SA in eq.2, although SA was not the highest correlated animal to the mean group direction (this was VI, id 13, with a value of 0.69). Thus, AL's movement appeared to be more sensitive to the movement of SA, rather than to the direction of the group as a whole.

5. **Discussion:**

Our results indicate that sparse data are able to capture movement patterns, and can be used to interrogate the individual movement behaviours that underpin the coordinated movement of social groups. Although we lost accuracy, as well as the power, to identify correct model structure as the data became increasingly sparse, this was ameliorated by increasing the extent of the data set. Given that increased observation extent will almost certainly result in animals encountering a broader array of social and ecological conditions, models fit to increasingly large extents sets should also provide more generalized fits, i.e., our approach is well suited to detecting common and consistent patterns of motion across contexts. This suggests that the use of force matching with sparse data is most suitable for questions concerning behavioural responses to frequently encountered social and ecological conditions (e.g., dyadic interaction within groups, common group formations/structures, or reducing isolation), but less useful for identifying specific behavioural responses to very rare events, where higher frequency data are necessary. Our simulation also suggested that parameter estimation of the contexts under which a behavioural switch might occur will similarly benefit from high resolution data. Importantly, however, comparisons to determine the relative magnitude of influential factors were less sensitive to data resolution (Fig. 5).

Given that force matching is derived from physics and deals with particle interactions, not the behaviour of animate agents, it raises an interesting issue regarding the interaction among animals, and what is meant by "force". We suggest using a less strict interpretation of force that does not assume that individual animals will always react consistently and instantaneously to all changes in their associates' positions. Rather, we should assume that individuals will influence others in a consistent but probabilistic fashion. For example, as the dominant male approaches an unrelated subadult male, we might predict that the subadult will move off. The distance moved and when movement is initiated, however, will depend on other external and internal factors. As such, it is not best modeled as an instantaneous reaction to the movement of the dominant male. The latter, at best, could be considered as the equivalent of the most minimal stimulus-response model of agent behaviour. In other words, we need to recognize that we are dealing with agents that may be attempting actively to control their local social and ecological environments and face certain trade-offs (e.g., safety from predation versus food density) that influence their responses accordingly. In this view, it is therefore possible that agents endogenously initiate their own movements in order to achieve a certain state (e.g., move in ways to ensure the presence of X number of animals in their field of view), rather than responding only to exogenous stimuli (i.e., movements are more 'response-stimulus' than stimulus-response, or more accurately, agents engage in an ongoing cycle of sensorimotor coordination with their socioecological environment: Barrett 2011).

Adopting this more dynamic, probabilistic view of action means that we should ask questions about the optimal spatial states an agent aims to achieve, and the actions needed to achieve such states, rather than focusing solely on the response to a particular group-mate spatial configuration. We could then develop the force matching approach to test between an array of hypothesized motion models that specify both the goal-states aimed for and the possible mechanisms by which these are achieved by comparing these to observed motion patterns. This would provide a means of identifying the complexity of the rules required to explain observed patterns of movement.

When we applied our modified version of the force matching approach to a focal individual in a baboon troop, we found patterns that suggested preferential movement towards certain individuals. Specifically, our focal animal was most sensitive to the highest-ranking females (rank 1=id 9, and rank 3=id 13), and the alpha male. These patterns accord with spatial and social association data collected independently of the GPS values (unpublished data): a topic we will explore in detail elsewhere. We observed a similar effect with the strong correlation between the direction to the highest ranking female and the mean direction of the group as a whole. By comparing both possibilities within one model, the force matching approach provides a tool that can help tease apart the relative influence of these different rules of thumb.

*5.1 Summary*

Our results suggest that, within limits, temporally sparse movement data can be used successfully to extract patterns of individual movement. We were able to demonstrate this by using simulation models to recover a pre-specified pattern of behaviour. Applying our approach to data from wild baboons, we were able to identify consistent behavioural responses in a focal baboon, suggesting that individual social interactions were largely responsible for group cohesion. Given the wide availability and low cost of handheld GPS devices, we suggest that sparse movement datasets of social groups can provide a valuable means for developing and empirically testing models of how individuals control their local social and ecological environments.

**Acknowledgements:**

We thank Dr. Parry Clark who collected the baboon spatial data, and Mel Lefebvre who provided helpful comments on the manuscript. The running of optimization algorithms were facilitated through the use of Westgrid/Compute Canada facilities. Cape Nature gave permission for the baboon research and funding for fieldwork was provided by Leakey Foundation (USA), NRF (South Africa) and NSERC (Canada) grants to SPH and LB. LB is also supported by NSERC's Canada Research Chairs Program (Tier 1). TB is supported by a FQRNT Postdoctoral Fellowship and the Canada Research Chairs program (LB).

22